\documentstyle[amsfonts,preprint,aps]{revtex}
\preprint{\bf Preprint UNO-HEP-98-06}
\tighten
\begin{document}
\draft
\title{
Theoretical Determination of the $\Delta N\gamma $
Electromagnetic Transition Amplitudes in the $\Delta (1232)$ Region
}
\author{
Milton Dean Slaughter \footnote{E-Mail address (Internet): mslaught@uno.edu \\Research Partially Supported by the National Science Foundation \\Abbreviated Version to be Submitted to Physical Review Letters}
}
\address{
Department of Physics, University of New Orleans, New Orleans, LA 70148
}
\date{September 1998}

\maketitle

\begin{abstract}
We utilize non-perturbative and fully relativistic methods to calculate
the\thinspace \thinspace $\Delta N\gamma $ electromagnetic transition
amplitudes $G_{M}^{*}(q^{2})$ (related to the magnetic dipole moment $%
M_{1^{+}}^{3/2}(q^{2})$), $G_{E}^{*}(q^{2})$ (related to the electric
quadrupole moment $E_{1^{+}}^{3/2}(q^{2})$), the electromagnetic ratio $%
R_{EM}(q^{2})\equiv
-G_{E}^{*}(q^{2})/G_{M}^{*}(q^{2})=E_{1^{+}}^{3/2}(q^{2})/M_{1^{+}}^{3/2}(q^{2})
$, and discuss their $q^{2}$ behavior in the $\Delta (1232)$ mass region.
These are very important quantities which arise in all viable quark, QCD, or
perturbative QCD models of pion electroproduction and photoproduction.
\end{abstract}


\newpage

\section{Introduction: The $\Delta N\protect\gamma $ Transition Form Factors
and Multipoles}

\label{i}

The $\Delta N\gamma $ transition form factors\cite{devenish77} $G_{M}^{\ast
}(q^{2})$, $G_{E}^{\ast }(q^{2})$, and $G_{C}^{\ast }(q^{2})$ and their
multipole counterparts $M_{1^{+}}$, $E_{1^{+}}$, and $S_{1^{+}}$ in nuclear
and elementary particle physics are very important in nuclear and elementary
particle physics because they provide a basis for testing theories of
effective quark forces or production models.

\begin{itemize}
\item  They are especially important in the understanding of perturbative
QCD (PQCD) models\cite{stoler93} involving gluon exchange mechanisms, tensor
interactions, or possible hybrid baryonic states;

\item  They are important in enhanced quark models in which the transition
form factors may be calculated as a function of $q^{2}$; in
electroproduction and photoproduction processes; in symmetry schemes such as
SU(6) and U(6,6), and Melosh transformations; in bag models; in dispersion
relation and Bethe-Salpeter approaches; in current algebra baryon sum rules;
and in nonperturbative methods such as lattice QCD, QCD sum rules, and
algebraic formulations.

\item  The fundamental reason that the transition form factors are such good
QCD probes lies in the fact that in many quark, symmetry, and potential
models, $G_{E}^{*}(q^{2})$ and/or $G_{C}^{*}(q^{2})$ are identically zero
thus giving rise to pure magnetic dipole $M_{1^{+}}$ transitions

\item  In the na\"{i}ve quark model, it can be shown that the quantity $%
E_{1^{+}}/M_{1^{+}}=-G_{E}^{*}/G_{M}^{*}\equiv $ ratio of the electric
quadrupole moment to the magnetic dipole moment $\;\equiv \;R_{EM}=\;0$ .
Experimentally, however, the $R_{EM}$ appears to be non-zero but small in
magnitude and of the order of a few percent. Most analyses predict the $%
R_{EM}$ to be small and negative at small momentum transfer, however a
recent analysis\cite{burkert} extracted the value $R_{EM}(q^{2}=-3.2$ $%
(GeV/c)^{2})\approx (+6\pm 2\pm 3)\%$. Subsequently, another even more
recent analysis\cite{beck}predicts that $R_{EM}=-(2.5\pm 0.2\pm 0.2)\%$ at
the maximum of the $\Delta (1232)$\ resonance and $R_{EM}=-3.5\%$ when
background scattering amplitude contributions are taken into consideration.

\item  Clearly, the capability of any particular theoretical model
(including ours) to predict accurately and precisely non-zero $R_{EM}$
values of the right sign and magnitude for particular values of $q^{2}$ in
agreement with experiment is critical. As has been noted, the $R_{EM}$ ratio
is especially effective for testing effective quark forces such as occur in
QCD one-gluon exchange tensor interactions, various types of enhanced quark
models, symmetry schemes such as $SU(6)$, $U(6,6)$, melosh transformations,
dispersion relations and sum rules (where the $\Delta $ always plays an
important role).
\end{itemize}

%%\newpage

\subsection{Importance of the ${\bf {\Delta \rightarrow N+\protect\gamma }}$
Transition Form Factors}

\begin{itemize}
\item  Provide a basis for testing theories of effective quark forces and
production models

\item  QCD: One gluon exchange mechanisms, tensor interactions, and possible
hybrid baryonic states.

\item  Enhanced Quark Models: They should be capable of predicting
accurately the $\Delta $--$N$ transition form factors as a function of $%
q^{2} $.

\item  Bag models of hadrons

\item  Current Algebra approaches to hadron physics

\item  Non-perturbative approaches to hadron physics such as lattice QCD

\item  Electroproduction and Photoproduction: Important for correct
theoretical description.

\item  Symmetry Schemes: The $\Delta $ always plays an important role in
models involving $SU(6)$, $U(6,6)$, etc., and melosh transformations.

\item  Dispersion relations: The $\Delta $ always plays an important role.

\item  Baryon Sum Rules: The $\Delta $ always plays an important role.
\end{itemize}

%%\newpage

\subsection{\ The ${\bf {\Delta \rightarrow N+\protect\gamma }}$ Transition
Form Factors and Transition Amplitudes}

In general one may write for the $\Delta \rightarrow N+\gamma $ transition
amplitude the following expression\cite{devenish77}:

\begin{equation}
\left\langle p(\vec{p},\lambda _{p})\right| j_{\mu }(0)\left| \Delta ^{+}(%
\vec{p}^{*},\lambda _{\Delta })\right\rangle =\frac{1}{{(}2\pi {)^{3}}}\sqrt{%
\frac{mm^{*}}{E_{p}E_{\Delta }}}\bar{u}_{P}(\vec{p},\lambda _{P})\left[
\Gamma _{\mu \beta }\right] u_{\Delta }^{\beta }(\vec{p}^{\,*},\lambda
_{\Delta })  \label{eq1}
\end{equation}

where

\begin{equation}
\begin{array}{ccl}
\Gamma _{\mu \beta } & = & i%
%TCIMACRO{\dfrac{3(m^{*}+m)}{2m} }%
%BeginExpansion
{\displaystyle{3(m^{*}+m) \over 2m}}%
%EndExpansion
(G_{M}^{*}-3G_{E}^{*})\Theta ^{-1}m^{*}q_{\beta }\epsilon _{\mu }(qp\gamma )
\\ 
& - & 
%TCIMACRO{\dfrac{3(m^{*}+m)}{2m} }%
%BeginExpansion
{\displaystyle{3(m^{*}+m) \over 2m}}%
%EndExpansion
(G_{M}^{*}+G_{E}^{*})\Theta ^{-1}[2\epsilon _{\beta \sigma }(p^{*}p)\epsilon
_{\mu }^{\;\sigma }(p^{*}p)\gamma _{5}-im^{*}q_{\beta }\epsilon _{\mu
}(qp\gamma )] \\ 
& + & 
%TCIMACRO{\dfrac{3(m^{*}+m)}{m} }%
%BeginExpansion
{\displaystyle{3(m^{*}+m) \over m}}%
%EndExpansion
G_{C}^{*}\Theta ^{-1}q_{\beta }[p\cdot qq_{\mu }-q^{2}p_{\mu }]\gamma _{5}.
\end{array}
\label{eq2}
\end{equation}

In Eqs. (\ref{eq1}) and (\ref{eq2}), the electromagnetic current is denoted
by $j_{\mu }$, $q\equiv p^{\ast }-p$, $p^{\ast }$ and $p$ are the
four-momenta of the $\Delta ^{+}$ and nucleon respectively. $\Theta
^{-1}\equiv \lbrack ((m^{\ast }+m)^{2}-q^{2})((m^{\ast }-m)^{2}-q^{2})]^{-1}$
is a kinematic factor which depends on $q^{2}$, $m^{\ast }$ (the $\Delta ^{+}
$ mass), and $m$ (the proton mass); $\lambda _{P}$ and $\lambda _{\Delta }$
are the helicities of the proton and $\Delta ^{+}$ respectively. We note
that the first, second, and third terms in Eq.(\ref{eq2}) induce transverse $%
\frac{1}{2}$ ($h_{3}$), transverse $\frac{3}{2}$ ($h_{2}$), and longitudinal
helicity transitions ($h_{1}$) respectively in the rest frame of the $\Delta
^{+}$ isobar\cite{devenish77}. $G_{M}^{\ast }$, $G_{E}^{\ast }$, and $%
G_{C}^{\ast }$ are related to the helicity form factors $h_{1}$, $h_{2}$,
and $h_{3}$ by the relations:

\begin{eqnarray}
h_{3} &=&-\frac{3(m^{*}+m)}{2m}(G_{M}^{*}-3G_{E}^{*})  \label{eq3} \\
h_{2} &=&-\frac{3(m^{*}+m)}{2m}(G_{M}^{*}+G_{E}^{*})  \nonumber \\
h_{1} &=&\frac{3(m^{*}+m)}{m}G_{C}^{*}.  \nonumber
\end{eqnarray}

%%\newpage

For the transition amplitude governing the virtual process $p\rightarrow
p+\gamma $, we have similarly

\begin{equation}
\left\langle p(\vec{s},\lambda )\right| j_{\mu }(0)\left| p(\vec{t},\lambda
^{*})\right\rangle =\frac{1}{{(2\pi )^{3}}}\sqrt{\frac{m{^{2}}}{E{_{\vec{p}}}%
E_{\vec{p}^{*}}}}\bar{u}_{p}({\vec{s}},\lambda )\left[ {\Gamma _{\mu }}%
\right] u_{p}(\vec{t},\lambda ^{*})  \label{eq4}
\end{equation}

where

\begin{equation}
\Gamma _{\mu }=[1-\widetilde{q}^{2}/(4m{^{2})}]^{-1}[(i/(4m^{2}))G_{M}(%
\tilde{q}^{2})\epsilon _{\mu }\left( \tilde{P}\tilde{q}\gamma \right) \gamma
_{5}+(1/(2m))G_{E}(\tilde{q}^{2})\tilde{P}_{\mu }],  \label{eq4.1}
\end{equation}

$\tilde{P}{_{\mu }}\equiv \widetilde{p}+\widetilde{p}^{*}$ , $\widetilde{q}=%
\widetilde{p}^{*}-\widetilde{p}$ with $\widetilde{p}^{*}=(\widetilde{p}^{*0},%
\vec{t})$ and $\widetilde{p}=(\widetilde{p}^{0},\vec{s})$. $G_{M}(\widetilde{%
q}^{2})$ and $G_{E}(\widetilde{q}^{2})$ in Eq.~(\ref{eq4.1}) are the
familiar nucleon Sachs form factors.

\subsection{Relationship Between the ${\bf \Delta N\protect\gamma }$\ Form
Factors and Multipoles}

The magnetic, electric, and coulombic multipole transition moments given by $%
M_{1^{+}}(q^{2})$, $E_{1^{+}}(q^{2})$, and $S_{1^{+}}(q^{2})$ can be written
\cite{cott} in terms of $G_{M}^{\ast }(q^{2})$, $G_{E}^{\ast }(q^{2})$, and $%
G_{C}^{\ast }(q^{2})$ . Indeed one has

\begin{eqnarray}
M_{1^{+}} &=&\alpha _{1}\;\sqrt{Q^{-}}\;G_{M}^{*}  \label{eq4.2} \\
E_{1^{+}} &=&\alpha _{2}\;\sqrt{Q^{-}}\;G_{E}^{*}  \nonumber \\
S_{1^{+}} &=&\alpha _{3}\;Q^{-}\sqrt{Q^{+}}\;G_{C}^{*}  \nonumber
\end{eqnarray}

where $\alpha _{1}$, $\alpha _{2}=-\alpha _{1}$, $\alpha _{3}$\ are
functions of parameters governing the process $\Gamma(\Delta \rightarrow \pi
N)$ (and in particular are dependent on the $\Delta $\ mass $m^{*}$) and
where $Q^{\pm }\equiv \sqrt{(m^{*}\pm m)^{2}-q^{2}}$.

%%\newpage

\subsection{${\bf {A_{\frac{1}{2}}}}$ and ${\bf {A_{\frac{3}{2}}}}$ Photon
Decay Helicity Amplitudes}

The total $\Delta $ radiative width $\equiv \Gamma _{\gamma }^{T}$, for
decay into $p+\gamma $ is given by:

\begin{equation}
\Gamma _{\gamma }^{T}=\frac{mq_{c}^{2}}{2m^{\ast }\pi }\sum_{\lambda =\frac{1%
}{2},\frac{3}{2}}A_{\lambda }^{2}  \label{eq5}
\end{equation}
where 
\begin{equation}
A_{\frac{1}{2}}=-e\left( \frac{\sqrt{3}}{8}\right) {\left[ \frac{{m^{\ast }}%
^{2}-m^{2}}{m^{3}}\right] }^{\frac{1}{2}}\left[ G_{M}^{\ast
}(0)-3G_{E}^{\ast }(0)\right] ,  \label{eq6}
\end{equation}
\vspace*{0.3in} $q_{c}^{2}=$ CM momentum, and 
\begin{equation}
A_{\frac{3}{2}}=-e\left( \frac{3}{8}\right) {\left[ \frac{{m^{\ast }}%
^{2}-m^{2}}{m^{3}}\right] }^{\frac{1}{2}}\left[ G_{M}^{\ast }(0)+G_{E}^{\ast
}(0)\right]   \label{eq7}
\end{equation}
\vspace*{0.5in} Experimentally\cite{pdg98},

\begin{equation}
A_{\frac{1}{2}}\cong (-140\pm 5)\times 10^{-3}GeV^{-1/2}\text{\qquad and
\qquad }A_{\frac{3}{2}}\cong (-258\pm 6)\times 10^{-3}GeV^{-1/2}
\label{eq7.1}
\end{equation}

%%\newpage

\section{Calculation}

\begin{itemize}
\item  Our treatment is non-perturbative and performed in a broken symmetry
hadronic world\cite{milt}. We do not require the use of ``mean'' mass
approximations;

\item  Physical masses are used at all times. Thus, $G_{E}^{*}$ is not
forced to equal zero as in the na\"{i}ve quark model;

\item  Our treatment is completely relativistic. Current conservation is
guaranteed. Additionally, the correct electromagnetic transition operator is
used in all calculations;

\item  We use the infinite-momentum frame for calculations or
equivalently--- ``infinite'' Lorentz boosts that are not always in the
z-direction, thus implicitly (and often explicitly) bringing into play
Wigner rotations resulting in mixed helicity particle states.

\item  In order to proceed with the calculation of $G_{M}^{*}(q^{2})$, and $%
G_{E}^{*}(q^{2})$, we consider helicity states with $\lambda =\pm {1/2}$ and 
$\lambda _{\Delta }=1/2$ (i.e. spin flip and non-flip sum rules) and the
non-strange ($S=0$) $L=0$ ground state baryons ($J^{PC}=\frac{1}{2}^{+},%
\frac{3}{2}^{+}$). We will ultimately find two independent constraint
equations which will then allow one to calculate $G_{M}^{*}(q^{2})$, and $%
G_{E}^{*}(q^{2})$, and their multipole counterparts.
\end{itemize}

%%\newpage \ 

\begin{itemize}
\item  It is well-known\cite{milt} that if one defines the axial-vector
matrix elements: $\,\left\langle p,1/2\right| A_{\pi ^{+}}\left|
n,1/2\right\rangle \equiv f=g_{A}(0)$, $\left\langle {\Delta ^{++},1/2}%
\right| A_{\pi ^{+}}\left| {\Delta ^{+},1/2}\right\rangle \equiv -\sqrt{{%
\frac{3}{2}}}\,g$, and $\left\langle {\Delta ^{++},1/2}\right| A_{\pi
^{+}}\left| {p,1/2}\right\rangle \equiv -\sqrt{6}\,h$, and applies
asymptotic level realization to the chiral $SU(2)\otimes SU(2)$ charge
algebra $\left[ {A_{\pi ^{+}},A_{\pi ^{-}}}\right] =2V_{3}$, then $h^{2}={%
(4/25)}f^{\;2}$ (the sign of $h=+(2/5)f$ , can be fixed by requiring that $%
G_{M}^{*}(0)>0$)and $g={(-}\sqrt{2}/5)f$.

\item  If one inserts the algebra $\left[ j_{3}^{\mu }(0){,}A_{\pi ^{+}}%
\right] =A_{\pi ^{+}}^{\mu }{(0)}$ ($j^{\mu }\equiv j_{3}^{\mu }+j_{S}^{\mu }
$ , where $j_{3}^{\mu }\equiv $ isovector part of $j^{\mu }$ and $j_{S}^{\mu
}$ is isoscalar) between the ground states $\left\langle B(\alpha ,\lambda
=\pm 1/2,\overrightarrow{s})\right| $and $\left| B^{\prime }(\alpha ,\lambda
=1/2,\overrightarrow{t})\right\rangle $ with $\left| \overrightarrow{s}%
\right| \rightarrow \infty ,\left| \overrightarrow{t}\right| \rightarrow
\infty $, where $\left\langle B(\alpha )\right| $ and $\left| B^{\prime
}(\beta )\right\rangle $ are the following $SU_{F}(2)$ related combinations: 
$\left\langle p,n\right\rangle $, $\left\langle p,\Delta ^{0}\right\rangle $%
, $\left\langle \Delta ^{++},p\right\rangle $, $\left\langle n,\Delta
^{-}\right\rangle $, $\left\langle \Delta ^{++},\Delta ^{+}\right\rangle $, $%
\left\langle {\Delta ^{+}},{\Delta ^{0}}\right\rangle $, $\left\langle
\Delta ^{0},\Delta ^{-}\right\rangle $ and $\left\langle \Delta
^{+},n\right\rangle $, then one obtains (we use $<N|\,j_{S}^{\mu }|\,\Delta
>=0$ ) the following two independent sum rule constraints:
\end{itemize}

\begin{center}
Spin Non-Flip Sum Rule
\end{center}

\begin{eqnarray}
<p,1/2,\vec{s}\,|j^{\mu }(0)|\,\Delta ^{+},1/2,\vec{t}> &=&\frac{5\sqrt{2}}{4%
}\left\{ -\frac{\left\langle p,1/2,\vec{s}\right| A_{\pi ^{+}}^{\mu
}(0)\left| n,1/2,\vec{t}\right\rangle }{2f}\right.  \label{eq8} \\
&&+\left. \left\langle p,1/2,\vec{s}\,|j_{3}^{\mu }(0)|\,p,1/2,\vec{t}%
\right\rangle \right\} .  \nonumber
\end{eqnarray}

and

\begin{center}
Spin Flip Sum Rule
\end{center}

\begin{equation}
<p,-1/2,\vec{s}\,|j^{\mu }(0)|\,\Delta ^{+},1/2,\vec{t}>=\frac{5}{8}\sqrt{2}%
\left\langle p,-1/2,\vec{s}\right| j_{3}^{\mu }(0)\left| p,1/2,\vec{t}%
\right\rangle  \label{eq9}
\end{equation}

%%\newpage

Now take the limit $\left| \overrightarrow{t}\right| \rightarrow \infty $
and $\left| \overrightarrow{s}\right| =\rightarrow \infty $ and evaluate
directly each of the matrix elements in Eq.(\ref{eq8}) and Eq. (\ref{eq9}).
We find respectively that:

\begin{equation}
G_{M}^{*}(q^{2})+\left[ \frac{2m^{*}(4m-m^{*})+m^{2}-2q^{2}}{%
2(m^{*2}+m^{2}-q^{2})}\right] G_{E}^{*}(q^{2})=  \label{eq10}
\end{equation}

\[
\frac{5\sqrt{3}}{3}\left[ \frac{2m^{*}m^{2}}{(m^{*}+m)(m^{*2}+m^{2}-q^{2})}%
\sqrt{\frac{(m^{*}+m)^{2}-q^{2}}{4m^{2}-q^{2}}}\right] G_{M}^{V}(q^{2}) 
\]

and

\begin{equation}
G_{M}^{*}(q^{2})-3G_{E}^{*}(q^{2})=\left[ \frac{5\sqrt{3}m\sqrt{-{\tilde{q}}%
^{2}}}{3(m^{*}+m)\left[ (m^{*}-m)^{2}-q^{2}\right] ^{1/2}}\right] G_{M}^{V}({%
\tilde{q}}^{2})\text{,}  \label{eq11}
\end{equation}

where in Eq. (\ref{eq11}), a collinear limit of $\left| \overrightarrow{t}%
\right| $ and $\left| \overrightarrow{s}\right| $ was taken in such a
fashion that $\left| \overrightarrow{s}\right| =r\left| \overrightarrow{t}%
\right| $ ($\overrightarrow{s}$ and $\overrightarrow{t}$ are taken along the
z-axis, $0<r\leq m^{2}/m^{*2}$) and ${\tilde{q}}^{2}$ and $q^{2}$ are
related by the equations ${\tilde{q}}^{2}=(1-r)\left[ q^{2}-(1-r)m^{*2}%
\right] $ and $q^{2}=\frac{(1-r)}{r}(m^{*^{2}}r-m^{2})$.

Eq. (\ref{eq10}) and Eq. (\ref{eq11}) may be solved for $G_{M}^{*}(q^{2})$
and $G_{E}^{*}(q^{2})$. The numerical results (Figure 1.) are as follows:

%%\newpage

\section{Results and Conclusions}

\begin{itemize}
\item  $G_{M}^{*}(q^{2})$, $G_{E}^{*}(q^{2})$, $R_{EM}(q^{2})$, $A_{\frac{1}{%
2}}$, and $A_{\frac{3}{2}}$ are computed in good agreement with experiment (Figure 1).

\begin{enumerate}
\item  At the $\Delta $ pole mass, we find that $G_{M}^{*}(0)=3.18$, $%
G_{E}^{*}(0)=+0.07$, and $R_{EM}(0)=-2.19\%$. We also determine that $A_{%
\frac{1}{2}}\cong -134\times 10^{-3}GeV^{-1/2}$, and $A_{\frac{3}{2}}\cong
-254\times 10^{-3}GeV^{-1/2}$.

\item  When the $\Delta $ mass is taken to be $1.232$ $GeV/c$, $%
G_{M}^{*}(0)=3.09$, $G_{E}^{*}(0)=+0.12$, and $R_{EM}(0)=-3.94\%$. We also
determine that $A_{\frac{1}{2}}\cong -128\times 10^{-3}GeV^{-1/2}$, and $A_{%
\frac{3}{2}}\cong -262\times 10^{-3}GeV^{-1/2}$;
\end{enumerate}

\item  We find that $R_{EM}(q^{2})$ is negative for $0\leq -q^{2}\lesssim
5\;GeV^{2}/c^{2}$, changes sign in the region $-q^{2}\approx
6-7\;GeV^{2}/c^{2}$, and very slowly approaches $1$ as $-q^{2}\rightarrow
\infty $;

\item  The PQCD prediction that $R_{EM}(q^{2})\rightarrow 1$ as $%
-q^{2}\rightarrow \infty $ is verified but only as an asymptotic condition
applicable only at very high momentum transfer\cite{stuart98};

\item  The $R_{EM}$ is particularly sensitive to the $\Delta $ mass (i.e.
mass parametrization used) when $0\leq -q^{2}\lesssim 1\;GeV^{2}/c^{2}$
(i.e. photoproduction).
\end{itemize}

%%\acknowledgments

%%This work was partially supported by the National Science Foundation.

%%\newpage

%%\begin{center}
%%{\large \bf REFERENCES }
%%\end{center}

\end{document}